\documentstyle[prb,aps,amsfonts,amssymb,floats,epsfig]{revtex}

\begin{document}

\draft
\twocolumn[\hsize\textwidth\columnwidth\hsize\csname @twocolumnfalse\endcsname

\title{Rung-singlet phase of the  $\bf{S}=\frac{1}{2}$ 
two-leg spin-ladder with four-spin cyclic exchange}

\author{K.P.~Schmidt$^1$, H. Monien$^2$ and G.S.~Uhrig$^1$}
\address{$^1$Institut f\"{u}r Theoretische Physik, Universit\"{a}t zu
  K\"{o}ln, Z\"{u}lpicher Stra{\ss}e 77, D-50937 K\"{o}ln, Germany}
\address{$2$Physikalisches Institut, Universit\"at Bonn, Nu{\ss}allee 12, 
D-53115 Bonn, Germany}

\date{September 25, 2002}

\maketitle
\begin{abstract}
 Using continuous unitary transformations (CUT) we calculate the one-triplet 
gap for the antiferromagnetic $S=\frac{1}{2}$ two-leg spin ladder with 
additional four-spin exchange interactions in a high order series expansion 
about the limit of isolated rungs. By applying a novel extrapolation technique
 we  calculate the transition line between the rung-singlet phase 
and a spontaneously dimerized phase with dimers on the legs. Using this 
efficient extrapolation technique we are able to analyze the crossover from 
strong rung coupling to weakly coupled chains.
\end{abstract}



\vskip2pc]
\narrowtext
After the discovery of high-$T_{\rm c}$ superconductivity in 1986, low 
dimensional quantum antiferromagnetism has attracted much attention in 
condensed matter physics. Recently it has become clear that the minimum 
magnetic model for cuprate systems has to include four-spin exchange terms 
besides the usual nearest neighbor Heisenberg exchange interaction 
\cite{honda93,loren99c,brehm99,matsu00a,colde01a,senth01,balen02,mulle02a,nunne02,schmi02b}. 
One important subclass of such models
are the two-leg ladder systems. The nearest-neighbor 
Heisenberg model on the two-leg ladder without four-spin interaction is a 
gapped spin liquid. This system is in the rung-singlet phase and first 
excitations are triplets \cite{shelt96,dagot92b,rice93,barne93}. In the 
limit of zero rung coupling there are two isolated gapless spin chains. 
Including four-spin exchange interactions several new quantum phases 
occur \cite{nerse97,lauch02b,hikih02}. Possible phases include
a spontaneously 
dimerized phase where the dimers are located in a meander-like structure on 
the legs, scalar and vector chirality phases, a region of dominant collinear 
spin and a ferromagnetic phase \cite{lauch02b}. However, real two-leg ladder
 cuprate systems are always in the rung-singlet phase but relatively close to 
the quantum phase transition to the spontaneously dimerized phase 
\cite{matsu00a,nunne02}. 
Therefore, it is in particular important to understand the 
properties of this transition.

In this paper we will calculate the gap around the limit of isolated rungs. 
We obtain reliable results in a wide range of parameters belonging to the
 rung-singlet 
phase. The transition curve to the spontaneously dimerized phase is computed. 
In addition, starting from the strong coupling limit of isolated rungs,
 the limit of isolated spin chains is discussed.
        
We consider the $S=\frac{1}{2}$ antiferromagnetic two-leg spin ladder plus 
additional four-spin exchange terms $H_{\rm cyc}$
\begin{mathletters}
\label{eq:Hamiltonian}
\begin{equation}
 H = J_\perp \sum\limits_i {\bf S}_{i,1}
{\bf S}_{i,2} + J_\parallel \sum_{i,\tau} {\bf S}_{i,\tau} {\bf S}_{i+1,\tau}
+H_{\rm cyc}
\end{equation}
where $i$ denotes the rungs and $\tau\in\{1,2\}$ the legs, and
\begin{eqnarray} 
\label{eq:Hcyc}
&&H_{\rm cyc} = 2J_{\rm cyc}\sum_{\rm plaquettes}[({\bf S_{1,i}S_{1,i+1}})
({\bf S_{2,i}S_{2,i+1}})\\
&& + ({\bf S_{1,i}S_{2,i}})({\bf S_{1,i+1}S_{2,i+1}}) - ({\bf S_{1,i}S_{2,i+1}})
({\bf S_{1,i+1}S_{2,i}})] \nonumber \ . 
\end{eqnarray}
\end{mathletters}

The exchange couplings along the rungs and along the legs are denoted by
$J_\perp$ and by $J_\parallel$, respectively. $J_{\rm cyc}$ denotes the 
strength of the four-spin magnetic exchange terms. There is also another way 
based on cyclic permutations $P_{ijkl}$ to include the
 leading four-spin exchange term. It differs in certain two-spin terms from 
Eq.\ (\ref{eq:Hamiltonian})
\begin{mathletters}
\label{eq:HamiltonianP}
\begin{eqnarray}
 H^{\rm p} &=& J_\perp^{\rm p} \sum\limits_i {\bf S}_{i,1}
{\bf S}_{i,2} + J_\parallel^{\rm p} \sum_{i,\tau} {\bf S}_{i,\tau} 
{\bf S}_{i+1,\tau}
+H_{\rm cyc}^{\rm p} \\
H_{\rm cyc}^{\rm p} &=&\frac{J_{\rm cyc}^{\rm p}}{2}\sum_{<ijkl>}\left( 
P_{ijkl}+P_{ijkl}^{-1}\right) \ .
\end{eqnarray}
\end{mathletters}

Both Hamiltonians are identical except for couplings along the diagonals 
\cite{brehm99}
if $J_\perp$ and $J_\parallel$ are suitably redefined.
First, we use Hamiltonian $H$ (\ref{eq:Hamiltonian}) since it is established 
that the four-spin terms are the most significant ones if the magnetic 
Hamiltonian is seen as effective model for the low-lying modes of a
realistic insulating three-band Hubbard model
\cite{mulle02a}. But results for Hamiltonian $H^{\rm p}$ 
(\ref{eq:HamiltonianP}) will also be presented.

We use a continuous unitary transformation (CUT)
 to map the Hamiltonian $H$ to an effective Hamiltonian 
$H_{\rm eff}$ which conserves the number of rung-triplets, i.e. 
$[H_0,H_{\rm eff}]=0$ where $H_0:=H|_{[J_\parallel=0,J_{\rm cyc}=0]}$ 
\cite{knett00a}. The ground state of $H_{\rm eff}$ is the rung-triplet vacuum.
 The effective Hamiltonian $H_{\rm eff}$ is calculated in order 11 in 
$x:=J_\parallel/J_\perp$ and $x_{\rm cyc}:=J_{\rm cyc}/J_\perp$. Thereby, we 
obtained the ground-state energy $E_0=\langle 0|H_{\rm eff}|0\rangle $ and the
 one-triplet dispersion $\omega(k)=\langle k|H_{\rm eff}|k\rangle - E_0$. The 
one-triplet dispersion $\omega(k)$ has a minimum for $k=\pi$, the one-triplet
 gap $\Delta(x,x_{\rm cyc}):=\omega(\pi)$.
By such perturbative approaches working on the operator level 
the spin ladder without cyclic exchange has been investigated previously
with great success \cite{trebs00,knett01b}.

The standard approach to calculate a phase transition line 
with series expansions is to use 
dlogPad\'{e} approximants on $\Delta(x,x_{\rm cyc})$. This yields reliable
 results only in a very small region about the exactly known phase transition 
point $[x=1/5,x_{\rm cyc}=1/5]$ (See grey square in Fig.\ \ref{fig3} or similarly
in Fig.\ \ref{fig4}). Generally, for 
$x=x_{\rm cyc}$ the dispersion and the gap are known exactly
\begin{eqnarray}
 \omega (k)/J_\perp &=& 1+(2\cos(k)-3)x \nonumber\\
 \Delta(x,x)/J_\perp &=& 1- 5x \quad .
\end{eqnarray}
The results extrapolated in $x$
 are reliable for $x\in [0.1,0.3]$ where the gap closes 
linearly in $x$ and $x_{\rm cyc}$. For Hamiltonian $H^{\rm p}$ 
(\ref{eq:HamiltonianP}) the analogous situation is found at 
and about the exact point $x^{\rm p}=x^{\rm p}_{\rm cyc}=1/4$ 
as shown in Ref.\ \onlinecite{mulle02b}. Note that
we use the parameters with superscript $^{\rm p}$ to distinguish
results for
the Hamiltonian $H^{\rm p}$ (\ref{eq:HamiltonianP})
clearly from those for the Hamiltonian $H$ (\ref{eq:Hamiltonian}).

In the following, we advance a recently introduced extrapolation 
technique \cite{schmi02a} in order to investigate the rung-singlet phase for 
larger/lower values of $x$ and $x_{\rm cyc}$. The main idea is to express the
 series expansion not in external parameters of the system like $x$ and 
$x_{\rm cyc}$, but in an \emph{internal} energy. 
Thereby, we combine high series 
expansion and renormalization group ideas. The natural \emph{internal}
 energy scale of the two-leg ladder is the one-triplet gap. In practice, 
we define the function 
\begin{equation}
  \label{G}
 G(x) = 1-\overline{\Delta}(x)=1-\frac{\Delta(x,rx)}{(1+x)J_\perp} 
\end{equation}
where  $r=x_{\rm cyc}/x = J_{\rm cyc}/J_\parallel$ 
will be kept constant for the extrapolation in $x$. 
The function $G(x)$ behaves like $G\propto x$ for $x\rightarrow 0$ so that any
 expansion in $x$ can be converted in an expansion in $G$. 
Using the expansion for $\Delta(x)$ we calculated the inverse function 
$x=x(G)$ as a series in $G$ up to order 11 from Eq.~(\ref{G}). The quantity 
$\overline{\Delta}=\Delta/[(1+x)J_\perp ]$ measures the gap in units of 
$J_\perp +J_\parallel$ to ensure empirically a monotonic behavior of 
$\overline{\Delta}$ as function of $x$. Then the existence of the inverse
 $x(G)$ is assured. Next we consider the derivative of $\overline{\Delta}(x)$
\begin{equation}
  \label{G2}
\frac{d\overline{\Delta}(x)}{dx} = -\frac{dG}{dx} \ .
\end{equation}   
Substituting $x=x(G)$ in Eq.~(\ref{G2}) we obtain 
\begin{equation}
 \label{G3}
 -\frac{dG}{dx} = P(G) \ ,
\end{equation} 
where $P(G)$ is the truncated series of order 10 in $G$. Note that even the
 convergence of the truncated series $P(G)$ is significantly better than the 
convergence of the truncated series $\Delta^\prime(x)$ in $x$ as
discussed in Ref.~\onlinecite{schmi02a}. 
Because the gap is a monotonic decreasing function for $r=$const we can use
 dlogPad\'{e} approximants for $P(G)$ since $-dG/dx$ is non-negative. 
Integrating Eq.~(\ref{G3}) yields
\begin{equation}
 \label{G4}
 -\int_0^{G_0} \frac{dG}{P(G)} = \int_{0}^{x_0} dx = x_0\quad .
\end{equation} 
Therefore, integrating the left hand side to $G_0=1$, i.e. $\Delta=0$, 
provides the phase transition point $[x_0,r x_0]$ for a given $r$. For any
 $G_0\in [0,1[$ the gap is $\Delta(x_0,rx_0)/J_\perp=(1+x_0)(1-G_0)$. In this
 way, $\Delta (x,x_{\rm cyc})$ is obtained.

\begin{figure}[t]
\centerline{\psfig{figure=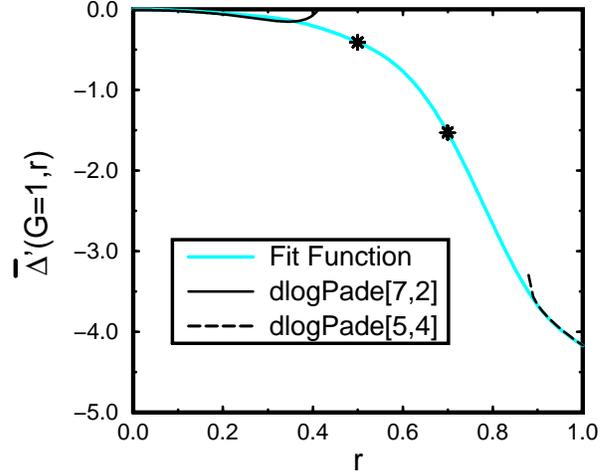,width=\columnwidth,clip=}}
\caption{For Hamiltonian $H$ (\ref{eq:Hamiltonian}); black lines show the 
dlogPad\'{e} approximants for 
$d\overline{\Delta}/dx$ at $G=1$ as a function of 
$r=x_{\rm cyc}/x$. The grey line is a fitted spline
which follows the asymptotic behavior Eq.~(\ref{smallr})
with $\lambda=0.41$ and $\lambda'=0.85$
at small values of $r$ and approximates  the available 
dlogPad\'{e} results. The points marked by stars are set by hand
to guide the spline smoothly in the intermediate region. The 
extrapolations in Figs.~3 and 4 require actually only the values 
$r\lessapprox 0.5$.}
\label{fig1}
\end{figure}
\begin{figure}[h]
\centerline{\psfig{figure=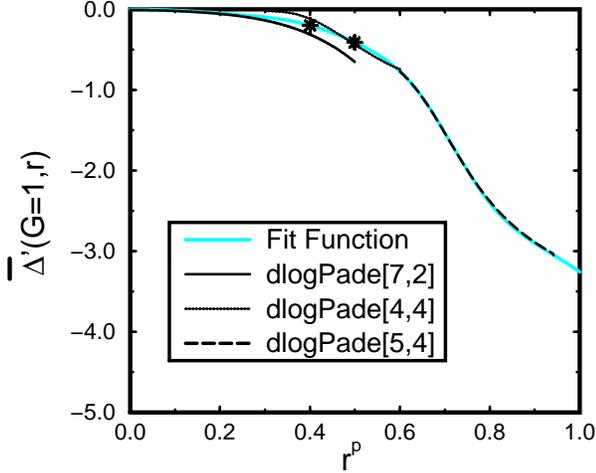,width=\columnwidth,clip=}}
\caption{Same as in Fig.~1 for Hamiltonian $H^{\rm p}$ (\ref{eq:HamiltonianP})}
\label{fig2}
\end{figure}

First, we examine the behavior of the gap in the limit of small $r$ and $G=1$.
 This corresponds to the situation of two spin chains which are weakly coupled
 by the four-spin interactions. Bosonization results show that the only 
relevant operator is the four-spin leg-leg interaction 
\cite{mulle02b}. The triplet gap scales as
\begin{equation}
 \label{Bos}
 \Delta= \lambda J_\perp-\lambda^\prime J_{\rm cyc}
\end{equation}   
in leading order in $J_\perp$ and $J_{\rm cyc}$. Here $\lambda$ and 
$\lambda^\prime$ are non-universal constants\cite{nerse97}. 
In our case we have a critical
 theory with central charge $c=\frac{3}{2}$ and SU(2) symmetry which is 
described as the $k=2$ Wess-Zumino-Witten model \cite{nerse97,hijii02}.
Rearranging Eq.~(\ref{Bos}) we obtain
\begin{equation}
  \label{Bos2}
  \frac{\Delta}{J_\parallel} = \frac{\lambda}{x_{\rm c}}\frac{x_{\rm c}-x}{x}
\end{equation}
where $x_{\rm c}=\lambda / (\lambda^\prime r)$ is the value of $x$ where the 
gap vanishes for given $r$. Therefore, the derivative of $\overline{\Delta}$ 
for small $r$ at $G=1$, i.e. $x=x_{\rm c}$, is given by
\begin{equation}
  \label{smallr}
  \overline{\Delta}^\prime (G=1,r) = 
-\frac{(\lambda^\prime r)^2}{\lambda+\lambda^\prime r}\quad .
\end{equation}
In the case of $r\rightarrow 0$ we expect $\overline{\Delta}^\prime=0$ and 
$\overline{\Delta}^\prime = -\lambda/x^2 = -\overline{\Delta}^2/\lambda$ from 
Eq.~(\ref{Bos2}). Exploiting 
$\overline{\Delta}^\prime = -\overline{\Delta}^2/\lambda$ in a biased 
dlogPad\'{e} approximant we find $\lambda=0.4\pm 0.03$ in very good agreement
 with Quantum Monte Carlo results 
$\lambda_{\rm QMC}= 0.41\pm 0.01$ from Ref.~\onlinecite{greve96}.

\begin{figure}[t]
\centerline{\psfig{figure=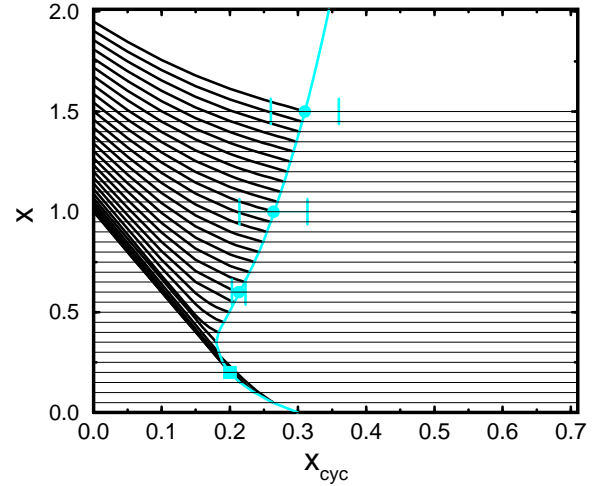,width=\columnwidth,clip=}}
\caption{Extrapolated spin gaps for Hamiltonian $H$(\ref{eq:Hamiltonian})
 in the $[x,x_{\rm cyc}]$-plane (see main text).
The  grey line is the obtained phase transition line $\Delta=0$ and the grey 
square is the exactly known transition point $[x=1/5,x_{\rm cyc}=1/5]$.
The points marked by grey circles and error bars indicate the estimated 
accuracy of the extrapolations. On the left
 side of the transition line the system is in the rung-singlet phase, on 
the right side in the spontaneously dimerized phase.
\label{fig3}}
\end{figure}
\begin{figure}[h]
\centerline{\psfig{figure=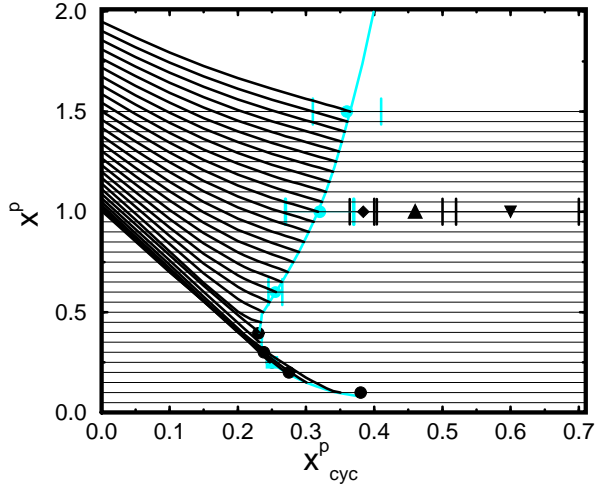,width=\columnwidth,clip=}}
\caption{Same as in Fig.~3 in the $[x^{\rm p},x^{\rm p}_{\rm cyc}]$-plane 
 for Hamiltonian $H^{\rm p}$ (\ref{eq:HamiltonianP}). The grey 
square is the exactly known transition point 
$[x^{\rm p}=1/4,x^{\rm p}_{\rm cyc}=1/4]$. The black circles are 
points taken from the curves in Ref.~\protect\onlinecite{mulle02b}.
The triangles are DMRG results (downward from  
Ref.~\protect\onlinecite{honda01};
upward from Ref.~\protect\onlinecite{lauch02b}). 
The diamond is determined from 
the maximization of the central charge by exact diagonalization
\protect\cite{hijii02}.
\label{fig4}}
\end{figure}

In Fig.~1 the solid line corresponds to the dlogPad\'{e} $[7,2]$ for 
$\overline{\Delta}^\prime (G=1,r)$. For $r < 0.3$ the asymptotic
formula (\ref{smallr}) is well reproduced by the approximant. 
A minute (not discernible) offset at $r=0$ occurs as a 
natural consequence of the dlogPad\'{e} approximation which describes 
a quantity of a given sign only. Using the value $\lambda=0.41 \pm 0.01$
we deduce for the second non-universal constant $\lambda'$ the value
\begin{equation}
\label{lamprim}
\lambda' = 0.85 \pm 0.2\ .
\end{equation}
If we perform the same analysis for Hamiltonian (\ref{eq:HamiltonianP})
we obtain Fig.~2 leading to the same result for $\lambda'$ given in 
Eq.~(\ref{lamprim}).
This fact corroborates the validity of the analysis and agrees perfectly with
 the
finding by M\"uller {\it et al.}\cite{mulle02b} stating that the relevant term
 in the cyclic exchange
is the leg-leg coupling so that both Hamiltonians 
(\ref{eq:Hamiltonian},\ref{eq:HamiltonianP})
lead to the same result for large leg couplings and small cyclic exchange 
couplings.

For larger values of $r$ or $r^{\rm p}$ we interpolate between various 
approximants.
This works better for Hamiltonian (\ref{eq:HamiltonianP}) (see Fig.\ 2)
 than for Hamiltonian (\ref{eq:Hamiltonian}) (see Fig.\ 1). But the 
interpolating
 functions are in any case quite similar. The uncertainty in the 
interpolation leads
 to the error bars in the subsequent extrapolations shown in Figs.\ 3 and 4.
 These extrapolations are done for values $r\lessapprox 0.5$
 by subtracting the interpolated values depicted 
 in Figs.\ 3 and 4 from the truncated series for $\overline{\Delta}^\prime (G)$
 so that we obtain the series of a function that vanishes at $G=1$. 
  We find that many in this way biased dlogPad\'{e}
 approximants yield reliable results. This supports our approach to include 
the properties of the weakly coupled chains in the extrapolations. 
Finally the subtracted bias is re-added to arrive at the proper result.

In the 
limit $x\rightarrow\infty$, we conclude from Eq.~(\ref{Bos2}) that the 
transition line converges against the asymptotic line
\begin{equation}
 x_{\rm cyc}^{\rm asympt} = {\lambda}/{\lambda^\prime} \approx 0.52 \pm 
0.14
\end{equation}
using the values for $\lambda$ and $\lambda^\prime$ obtained above. This result
holds again for both Hamiltonians  
(\ref{eq:Hamiltonian},\ref{eq:HamiltonianP}).
We cannot confirm the value of $x_{\rm cyc}^{\rm asympt} =0.22$ advocated in
Ref.\ \onlinecite{mulle02b}.

In Fig.~3 the extrapolated values of the spin gap of the Hamiltonian 
(\ref{eq:Hamiltonian}) in the $[x,x_{\rm cyc}]$-plane are presented. 
The black solid lines denote 
$\Delta(x_0,x_{\rm cyc})$ for a fixed $x_0$ as a function of $x_{\rm cyc}$.
 These lines are shifted by $x_0$ in $x$-direction producing a quasi 
three-dimensional plot. The end-point of a black line corresponds to 
$\Delta (x,x_{\rm cyc})=0$. These points yield the grey solid line which is 
the transition line between the rung-singlet phase and the spontaneously 
dimerized phase. As discussed above, we use biased approximants in the range 
$x\in[0.3,\infty[$ for the transition line. In the range $x\in[0.1,0.3]$ the 
unbiased extrapolations are safe due to the good convergence of the series 
obtained near the exactly known transition point (grey square). In the limit
 $x\rightarrow 0$ even the truncated series gives quantitative results. Using 
Eq.~(\ref{G4}) one finds in addition strong evidence for 
\begin{equation}
 \frac{d\overline{\Delta}}{dx} \propto (1-G)^{\eta} 
\end{equation}
at $x=0$ where $\eta=0.3\pm 0.02$. The transition point, i.e. $\Delta=0$, for 
$x=0$ is found to be $[0,0.3\pm 0.002]$.

The smooth connection between the different extrapolations corroborates the 
reliability of our results in a wide region in the $[x,x_{\rm cyc}]$-plane.

In Fig.~4 the corresponding results for spin gap of the Hamiltonian 
\ref{eq:HamiltonianP} in the  $[x^{\rm p},x^{\rm p}_{\rm cyc}]$-plane 
are depicted. The biased extrapolation is used for $x^{\rm p} \gtrapprox0.4$.
Besides quantitative differences there occurs one qualitative difference
at low values of $x^{\rm p}$. For Hamiltonian (\ref{eq:HamiltonianP})
no closing of the gap on increasing $x^{\rm p}_{\rm cyc}$ for $x^{\rm p}\lessapprox 0.1$
was found in agreement with the results in Ref.\ \onlinecite{mulle02b}.
This is the reason why the grey line is not prolonged below $x^{\rm p}=0.1$.
Apart from this point, the shape of the transition line is similar for both 
Hamiltonians. 

Quantitatively, it is interesting to compare to results obtained
by other approaches, see Fig.~4. Refs.\ \onlinecite{honda01,lauch02b}
 use density matrix renormalization. Another work analyzes the 
finite size scaling to determine the maximum central charge $c$. The cyclic
exchange at which it appears determines  the critical 
$x_{\rm cyc}^{\rm p}$, see Ref.~\onlinecite{hijii02}. 
The spread and the error bars of the various results allow to assess 
the accuracy of the data. We conclude that it is at present not yet
settled where precisely the transition between gapped rung-singlet and
dimerized phase occurs for the isotropic ladder. We propose to carry out
careful finite-size scaling on data for \emph{periodic} systems to clarify this
issue. Note that the symmetry change between the rung-singlet and the dimerized
phase cannot be represented properly in a {\em single} open system.

In summary, we have investigated the rung-singlet phase of the $S=\frac{1}{2}$ 
two-leg spin ladder with additional four-spin interactions. We used a 
continuous unitary transformation to calculate the one-triplet gap in a high 
order series expansion about the limit of isolated rungs. The use of an 
internal energy scale as the new expansion variable enabled us to calculate 
the transition line between the rung-singlet phase and the spontaneously 
dimerized phase reliably in a wide region of parameter space. 
Our results are consistent with the  field theoretic results in the limit 
of weakly coupled chains. We reproduce properties of the
bosonization results in that limit. In addition, we 
 give an estimate of the non-universal constants $\lambda$ and 
$\lambda^\prime$ which appear in these bosonization treatments. The value for
 $\lambda$ is in very good agreement with quantum Monte Carlo results\cite{greve96}. We 
have given an example that  the combination of high order series expansion and 
renormalization group ideas can be a powerful tool.
        
We thank A.~B\"uhler, C.~Knetter, U.~L\"ow and E.~M\"{u}ller-Hartmann for
helpful 
discussions and the DFG for financial support in SP 1073 and in SFB 608.


\end{document}